\documentclass[11pt]{article}
\usepackage{blois,epsfig}

\def\be{\begin{equation}}
\def\ee{\end{equation}}
\def\bea{\begin{eqnarray}}
\def\eea{\end{eqnarray}}

\def\la{Lyman-$\alpha$\,}
\begin{document}
\vspace*{4cm}
\title{COSMOLOGY WITH THE LYMAN-$\alpha$ FOREST IN THE WMAP ERA}

\author{MATTEO VIEL}

\address{Institute of Astronomy, Madingley Road, CB3 0HA Cambridge, UK}

\maketitle\abstracts{In the WMAP era of high precision cosmology an accurate
determination of the matter power spectrum from \la forest data becomes
crucial. When combining the matter power spectrum derived from CMB
experiments with that
inferred from \la absorption an evidence for a running spectral
index and a primordial index $n < 1$ arises (Verde et al. 2003). 
In this talk, I will describe some results obtained from
a sample of 27 high resolution and high signal-to-noise quasar spectra
(the LUQAS sample) and I will address possible systematic errors that can
affect the estimate of the flux power spectrum.}

\section{Introduction}
The \la forest offers a unique probe of our Universe at redshift and
scales not probed by any other observable.  Thus, the use of absorption
spectra to probe the dark matter power spectrum has been widely
investigated by a large number of authors (e.g. Croft et
al. \cite{croft}; Gnedin \& Hamilton \cite{gnedin}; Zaldarriaga et
al. \cite{zaldarriaga}).  At these scales the matter power spectrum is
sensitive to possible cut-off expected if the dark matter were warm
dark matter, can give constraints on the matter fraction in neutrinos
and allows to investigate the gravitational growth of structure and
possibly the redshift evolution of dark energy (Viel et
al. \cite{viel1}; Mandelbaum et al. \cite{mandelbaum}; Lidz et
al. \cite{lidz}).

Croft et al.  \cite{croft} found that the power spectrum inferred from
\la forest is consistent with a $\Lambda$CDM model ($\Omega_{\Lambda}=0.6,
\Omega_{M}=0.4, h = 0.65$) with $n=0.93$ and
$\sigma_8=0.7$. Hui et al.  \cite{hui} made a very detailed study of
possible systematic effects which can occur in this estimate. Verde et al.  \cite{verde} combined Croft et
al. data points with WMAP results and concluded that there is evidence for a running spectral
index and for a tilt in the primordial power spectrum. However, a
recent analysis made by Seljak et al.  \cite{seljak}, who argued for a
larger error bar and for a smaller value
of the effective optical depth, showed that there is no evidence for a running
spectral index nor for a tilt in the power spectrum.  An accurate
determination of the power spectrum is thereby crucial especially now
that with new data set coming out a precision of the order of few
percent could be achieved (Mandelbaum et al. \cite{mandelbaum}). In
this talk, I will describe some systematics errors which can
significantly affect the estimate of the flux power spectrum.

\section{The LUQAS sample: systematic errors in the flux power spectrum}
The LUQAS sample \footnote{\rm www.ast.cam.ac.uk/$\sim$rtnigm/luqas.html}
(Large Uves Quasar Absorption Spectra) consists of 27 QSOs. The median
redshift of the sample is $z=2.25$ and the total redshift path is
$\Delta z=13.75$. The typical signal-to-noise ratio is $\sim 50$ and
the pixel size is 0.05 \AA. For the data reduction and a more
extensive description of the sample we refer to Kim et al.
\cite{kim}.  In the left panel of Figure \ref{fig:continuum} we show
the 3D flux power spectrum from a subsample of the LUQAS sample at a
median redshift $z=2.04$ (diamonds) and we compare it with the results
of Croft et al. 2002 (triangles). There is good agreement between the
two estimates over a wide range of wave-numbers. In particular, we
find the same fluctuation amplitude at $k \sim 0.03$ s/km and a shallower
slope at large scales $k<0.03$ s/km than Croft et al. (2002) but
consistent within the errors.

\subsection{Continuum fitting errors and metal lines}
\begin{figure}
\psfig{figure=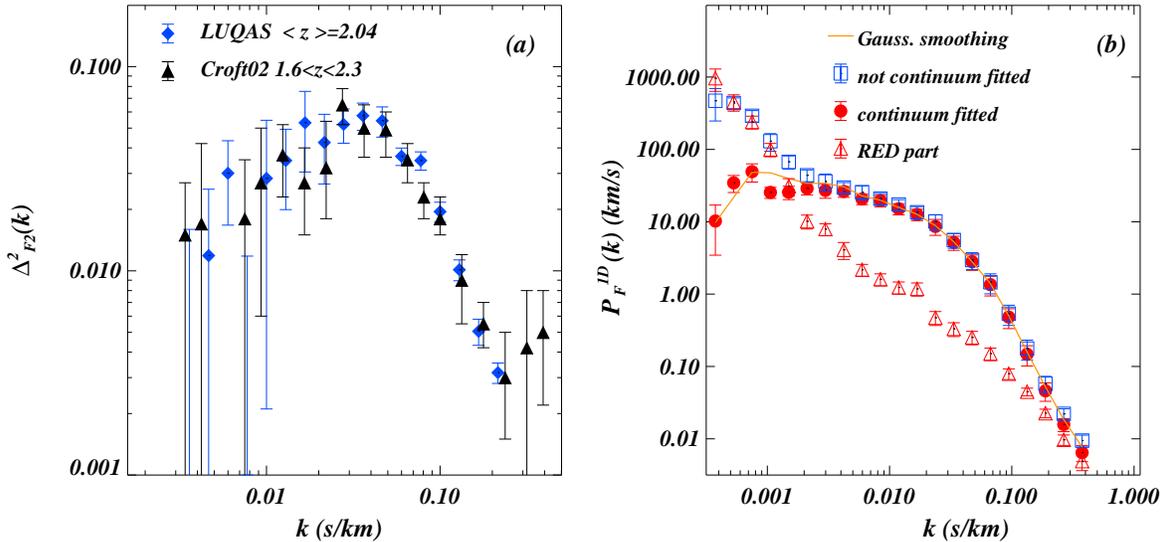,height=8.0cm,width=16cm}
\caption{{\underline {Left}:} 3D flux power spectrum plotted in adimensional
units from the LUQAS sample at $<z>=2.04$ (diamonds) and from Croft et
al. (2002, triangles). {\underline {Right}:} Effect of continuum fitting on the
1D flux power spectrum for different flux estimators: the fitted
spectra (red circles); the not continuum fitted spectra (squares); the
not continuum fitted spectra smoothed with a Gaussian filter with
$\sigma=25$ \AA (continuous line); the part of the flux spectrum
redward of \la emission (triangles). Error bars are
jack-knife estimates.
\label{fig:continuum}}
\end{figure}

In the right panel of Figure \ref{fig:continuum} we show the effect of
continuum fitting on the flux power spectrum. Red circles show the one
dimensional flux power spectrum computed from the {\it continuum fitted}
spectra, while the empty squares have been obtained from the {\it not
continuum fitted} spectra. There is more power for the latter at
$k<0.003$ s/km: this is where the continuum fluctuations from the
distant source start to dominate. In Figure 1 we plot the flux power
spectrum redward of \la emission (1265 \AA - 1393\AA) which is
consistent with the other two for $k<0.003$ s/km, supporting the
conclusion that at these scales and for these spectra the power is
continuum dominated. In the same Figure the continuous line represents
the power spectrum of the observed flux obtained by a local estimate of the
mean flux with a Gaussian filter with  $\sigma=25$\AA: the result is
consistent with the fitted flux power spectrum. This indicates that
simple way of continuum fitting the spectra, such as a Gaussian
filtering to obtain a local estimate of the mean flux, are promising. In
Kim et al. \cite{kim} we quantify the effect of metal lines on the flux
power spectrum which is less than 10\% at scales $k<0.01$ s/km and
rises up to 50\% at smaller scales. This estimate has been obtained from a
subset of 13 quasar spectra of the LUQAS sample for which we have the
metal lines list.

\subsection{Strong absorption systems}
\begin{figure}
\psfig{figure=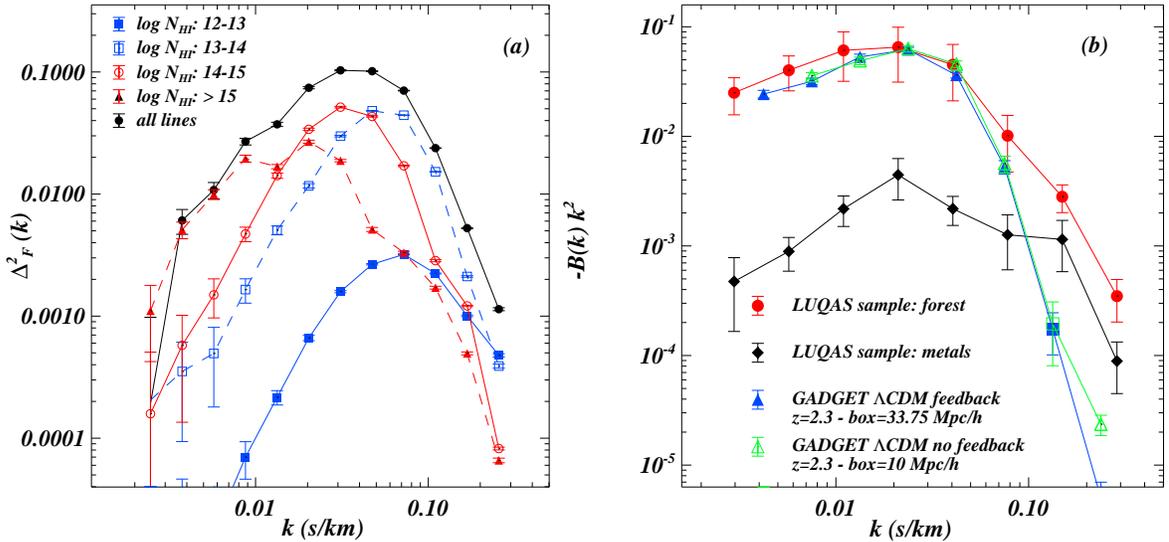,height=8.0cm,width=16cm}
\caption{{\underline {Left}:} Contribution of absorbers in different
column density ranges to the 3D flux `random' power
spectrum. {\underline {Right}:} The flux bispectrum: the LUQAS flux
bispectrum in the range $2<z<2.4$ (filled circles); bispectrum of the
metal lines (diamonds); bispectrum from hydro-dynamical simulations
with different box sizes and with and without feedback (filled and
empty triangles, respectively).
\label{fig:strong}}
\end{figure}

In the left panel of Figure \ref{fig:strong} we show the contribution
to the 3D flux power spectrum from absorbers with different column
densities (Viel et al. \cite{viel2}). From a subset of 8
spectra of the LUQAS sample fitted with VPFIT we create a new sample
of 300 `random' spectra for which the positions of the lines have been
randomly shifted along the spectrum. We split the contribution of
different absorbers by selecting only the lines in a given column
density range when creating the new set of spectra. The strongest contribution
comes from lines $13.5 < \log (N_{HI}/cm^{-2}) < 14.5$. These are
lines where the ``curve of growth'' which describes the relation
between equivalent width and column density changes from the linear to the flat regime due to saturation.
However, the contribution of systems with $\log N_{HI} > 14$ at
wavenumbers $k<0.01$ s/km is significant and dominates at large
scales. In addition, the contribution of these systems to the mean flux
decrement is of the order of 20\% (Viel et al. \cite{viel2}).

This will have profound implications for attempts to use numerical
simulations together with quasar absorption spectra to infer amplitude
and slope of the dark matter power spectrum with high accuracy. In
fact, numerical simulations of the \la forest often underpredict the
number of strong absorption systems. The calibration of numerical
simulations which underreproduce observed strong absorption is likely
to underestimate the inferred {\it rms} fluctuation amplitude and the
slope of the primordial dark matter power spectrum. In a combined
analysis with other data which constrain the dark matter power
spectrum on large scales, this can result in a spurious detection of a
running spectral index.

\section{The flux bispectrum}
In the right panel of Figure \ref{fig:strong} we show the one
dimensional flux bispectrum (Viel et al. \cite{viel3}). The
circles have been obtained from the LUQAS sample ($2<z<2.4$), the
diamonds represent the metal lines contribution which is significant
at $k>0.1$ s/km. Typical errors on the observed bispectrum as obtained from a jack-knife estimate are of the order of 50\%.

In this Figure the flux bispectrum from two hydro-dynamical
simulations with feedback (full triangles) and without feedback (empty
triangles) is reported. There is substantial agreement between these
two estimates (including galactic feedback changes the bispectrum by
less than 10\%), showing that feedback in the form of strong winds
from star forming galaxies does not affect this statistic (see Theuns
et al. \cite{theuns} for other statistics). This supports
the idea that the volume filling factor of the feedback regions is
small and thereby the \la forest is an unbiased probe of the Universe
at these redshifts. In the range $0.007$ s/km $<k
<0.07$ s/km there is agreement between hydro simulations and data,
while the discrepancy at smaller scales is due to the presence of
metal lines. An analytical prediction based on second order
perturbation theory in the framework of the fluctuating Gunn-Peterson
approximation is also in rough agreement with the data.   For the LUQAS sample the error bars are too large
to discriminate between models with very different 3D distribution of
\la absorption. In fact, we found that the bispectrum computed from a
set of randomized absorption spectra, for which a shift in wavelength
has been added to absorption lines identified with VPFIT, is in
agreement with the observed one. If it were possible to reduce these
error bars with a larger sample the bispectrum can become an important
tool for probing the growth of gravitational structures in the
Universe at $z>2$ (Viel et al. \cite{viel3}).

\section{Conclusions}
In this talk I presented some results from the LUQAS sample which
consists of 27 high resolution quasar spectra at a median redshift
$z=2.25$. The main conclusions are here summarized: {\it i)} the flux
power spectrum is consistent with the results found by Croft et
al. (2002); {\it ii)} the continuum fitting errors affect the power
spectrum at $k<0.003$ s/km; {\it iii)} strong absorptions systems have
been found to contribute significantly (up to 50 \%) to the flux power spectrum at
large scales and to the mean flux decrement ($\sim 20\%$); {\it iv)} the flux
bispectrum is a promising and robust statistics which needs to be
further investigated with a larger sample.

\section*{Acknowledgments}
I thank my collaborators: R. Carswell,  S. Cristiani, M. Haehnelt,
A. Heavens, L. Hernquist,  S. Kay, T.-S. Kim, S. Matarrese, D. Munshi, J. Schaye,
V. Springel,  T. Theuns, P. Tzanavaris, Y. Wang. I thank D. Weinberg,
R. Croft and L. Hui for
useful discussions. This work is supported by the European Community
Research and Training Network ``The Physics of the Intergalactic
Medium'' and by PPARC.


\section*{References}
\begin{thebibliography}{99}

\bibitem{croft} Croft R.A.C., Weinberg D. H., Bolte M., Burles S., Hernquist L., Katz N.,Kirkman D., Tytler D., 2002, ApJ, 581, 20 

\bibitem{gnedin} Gnedin N.Y., Hamilton A.J.S., 2002, MNRAS, 334, 107

\bibitem{hui} Hui L., Burles S., Seljak U., Rutledge R. E., Magnier E.,
Tytler D., 2001, ApJ, 552, 15

\bibitem{kim} Kim T.-S., Viel M., Haehnelt M.G., Carswell R.F.,
Cristiani S., 2003, MNRAS, in press

\bibitem{lidz} Lidz A.,  Hui L., Crotts P.A., Zaldarriaga M., 2003, submitted,  astro-ph/0309204

\bibitem{mandelbaum} Mandelbaum R., McDonald P., Seljak U., Cen R.,
2003, MNRAS, 344, 776

\bibitem{seljak} Seljak U., McDonald P., Makarov A., 2003, MNRAS, 342, 79

\bibitem{theuns} Theuns T., Viel M., Kay S., Schaye J., Carswell R.F., Tzanavaris P., 2002, ApJ, 578, L5

\bibitem{verde} Verde et al., 2003, ApJS, 148, 195

\bibitem{viel1} Viel M., Matarrese S., Theuns T., Munshi D., Wang Y.,
2003, 340, L47

\bibitem{viel2} Viel M., Haehnelt M.G., Carswell R.F., Kim T.-S., 2003,
submitted to MNRAS, astro-ph/0308078  

\bibitem{viel3} Viel M., Matarrese S., Heavens A., Haehnelt M.G., Kim
T.-S., Springel V., Hernquist L., 2003, astro-ph/0308151

\bibitem{zaldarriaga} Zaldarriaga M., Scoccimarro R., Hui L., 2003,
ApJ, 584, 559

\end{thebibliography}
\end{document}